\documentclass[11pt,preprint]{aastex}

\newcommand{\EffP}{\epsilon(>p)}
\newcommand{\EffOneFour}{\epsilon_{1.4}(>m_p c)}

\newcommand{\EffTen}{\epsilon_{10}(>10 m_p c)}

\newcommand{\ave}[1]{<\!\!{#1}\!\!>}
\newcommand\too{\, \rightarrow \, }
\newcommand{\Nmin}{N_{\mathrm{min}}}
\newcommand{\rMC}{r_{\mathrm{MC}}}

\newcommand\rgz{r_{g,0}}

\newcommand{\thetaSK}{\theta_{\mathrm{sk}}}
\newcommand{\deltime}{\delta t}
\newcommand{\deltheta}{\delta \theta}
\newcommand{\delmax}{\delta \theta_{\mathrm{max}}}

\newcommand{\fracinj}{f_{\mathrm{inj}}}
\newcommand{\GamNL}{\Gamma_{\mathrm{NL}}}
\newcommand{\GamTP}{\Gamma_{\mathrm{UM}}}
\newcommand{\gamux}{\gamma_u(x)}
\newcommand{\UDUave}{\ave{p_f/p_i}_{\mathrm{u-d-u}}}
\newcommand{\DUDave}{\ave{p_f/p_i}_{\mathrm{d-u-d}}}

\newcommand{\ppf}{p_{\mathrm{pf}}}
\newcommand{\Pret}{{\cal P_{\mathrm{ret}}}}
\newcommand{\rel}{relativistic}
\newcommand{\nonrel}{nonrelativistic}
\newcommand{\ultrarel}{ultrarelativistic}
\newcommand{\transrel}{trans-relativistic}

\newcommand{\gamsk}{\gamma_0}

\newcommand{\gaminj}{\gamma_{\mathrm{inj}}}

\newcommand{\mc}{Monte Carlo}

\newcount\Inum
\newcount\IInum
\newcount\IIInum
\newcount\IVnum
\def\I{\global\multiply\IInum by 0 \global\multiply\IIInum by 0
            \global\multiply\IVnum by 0 \global\advance \Inum by 1
            {\the\Inum. }}
\def\II{\global\multiply\IIInum by 0\global\multiply\IVnum by 0
       \global\advance \IInum by 1 {\the\Inum.\the\IInum. }}
\def\III{\global\multiply\IVnum by 0\global\advance \IIInum by 1
            {\the\Inum.\the\IInum.\the\IIInum. }}
\def\IV{\global\advance \IVnum by 1
            {\the\IVnum. }}
\newcount\listnorom
\newcommand\listromanDE{\global\advance \listnorom by 1 
{\lowercase\expandafter{\ (\romannumeral\listnorom)}\ }}

\newcommand{\xx}[1]{\!\times\!10^{#1}}

\newcommand\itt{ }
\newcommand\bff{ }

\newcommand\TP{test-particle}

\newcommand\kmps{km s$^{-1}$}

\newcommand\Vsk{u_0}
\newcommand\vinj{v_\mathrm{inj}}
\newcommand\Einj{E_\mathrm{inj}}

\newcommand\pmax{p_\mathrm{max}}

\newcommand\etal{et al.}

\newcommand\alf{Alfv\'en}

\shorttitle{}
\shortauthors{}

\begin{document}     
   
\title{Nonlinear Particle Acceleration in Relativistic Shocks}

\author{
Donald C.~Ellison\altaffilmark{1} 
and Glen P. Double\altaffilmark{1}}

\altaffiltext{1}{Department of Physics, North Carolina State
 University, Box 8202, Raleigh NC 27695, U.S.A.; E-mail:
 don\_ellison@ncsu.edu; gpdouble@unity.ncsu.edu; 919-515-7227}

\slugcomment{{\it Astroparticle Physics}, in press, April 2002}

\begin{abstract}
Monte Carlo techniques are used to model nonlinear particle
acceleration in parallel collisionless shocks of various speeds,
including mildly relativistic ones. 
When the acceleration is efficient, the backreaction of accelerated
particles modifies the shock structure and causes the compression
ratio, $r$, to increase above test-particle values.  Modified shocks
with Lorentz factors, $\gamsk \lesssim 3$, can have compression ratios
considerably greater than $3$ and the momentum distribution of
energetic particles no longer follows a power law relation.  These
results may be important for the interpretation of gamma-ray bursts if
mildly \rel\ internal and/or afterglow shocks play an important role 
accelerating particles that produce the observed radiation.
For $\gamsk \gtrsim 10$, $r$ approaches $3$ and the so-called
`universal' test-particle result of $N(E) \propto E^{-2.3}$ is
obtained for sufficiently energetic particles. In all cases, the
absolute normalization of the particle distribution follows directly
from our model assumptions and is explicitly determined.
\end{abstract}


\section{Introduction}

Most collisionless shocks in astrophysics are nonrelativistic, i.e.,
the flow speed of the unshocked plasma in the reference frame at rest
with the shock, $u_0$, is much less than the speed of light,
$c$. Particle acceleration in such shocks has been studied extensively
in both the linear and nonlinear regimes, and we
refer the reader to reviews which describe the basic features of the
shocks, the energetic particles they produce, and the many
applications \citep[e.g.,][]{Drury83,BE87,JE91}.
Relativistic shocks, where the flow speed Lorentz factor $\gamsk = [1
- (u_0/c)^2]^{-1/2}$ is greater than a few, are likely to be much less
common than nonrelativistic ones, but may occur in extreme objects
such as pulsar winds, hot spots in radio galaxies, and gamma-ray
bursts (GRBs).  Largely motivated by the application to GRBs,
relativistic shocks have recently received considerable attention by a
number of researchers \citep[e.g.,][]{BedOstrow96,KGGA2000,AGKG2001}.
However, except for some preliminary work done over a decade ago
\citep[][]{SK87,EllisonJapan91,EllisonPoland91}, current descriptions
of relativistic shocks undergoing first-order Fermi acceleration are
test particle approximations that do not include the backreaction of
the accelerated particles on the shock structure.
This may be a serious limitation of \rel\ shock theory in
applications, such as GRBs, where high particle acceleration
efficiencies are often assumed.  Here, we present results for
nonlinear acceleration where the backreaction of the accelerated
population on the \rel\ shock structure is included self-consistently.

In collisionless shocks, charged particles gain energy by scattering
back and forth between the converging upstream and downstream plasmas.
This basic physical process, called diffusive or first-order Fermi
shock acceleration, is the same in \rel\ and \nonrel\ shocks.
Differences in the mathematical description and outcome of the process
occur, however, because energetic particle distributions are nearly
isotropic in the shock reference frame in \nonrel\ shocks (where $v
\gg u_0$; $v$ is the individual particle speed), but are highly
anisotropic in \rel\ shocks (since $v \sim u_0 \sim c$)
\citep[e.g.,][]{Peacock81}.
The description of particle diffusion and energy gain is far more
difficult when $\gamsk \gg 1$ because the diffusion approximation,
which requires nearly isotropic distribution functions, cannot be
made. Because of this, \mc\ simulations, where particle scattering and
transport are treated explicitly, and which, in effect, solve the
Boltzmann equation with collective scattering
\citep[e.g.,][]{EE84,KS87b,EJR90,ER91,Ostrow91,BedOstrow96}, offer
advantages over analytic methods.  This is true in the test-particle
approximation, where analytic results exist, but is even more
important for nonlinear \rel\ shocks.

In \nonrel\ shocks, for $v \gg u_0$, a diffusion-convection equation
can be solved directly 
for infinite, plane shocks
\citep[e.g.,][]{ALS77,BO78}, yielding the
well-known result
\begin{equation}
\label{eq:TPpowerlaw}
f(p) \, d^3 p\propto
p^{-\sigma} \, d^3p
\quad \hbox{with} \quad
\sigma = 3 r/ (r-1)
\ ,
\end{equation}
where $r$ is the shock compression ratio, $p$ is the momentum, and
$f(p)\, d^3 p$ is the number density of particles in $d^3p$.
Eq.~(\ref{eq:TPpowerlaw}) is a steady-state, test-particle result
with an undetermined normalization, but the spectral index, $\sigma$,
in this limit is independent of the shock speed, $u_0$, or any details
of the scattering process as long as there is enough scattering to
maintain isotropy in the local frame. To obtain an absolute injection
efficiency, or to self-consistently describe the nonlinear
backreaction of accelerated particles on the shock structure (at least
when the seed particles for acceleration are not fully \rel\ to begin
with), techniques which do not require $v \gg \Vsk$ must be
used. Furthermore, 
for particles that do not obey $v \gg \Vsk$ 
additional assumptions must be made for how
these particles 
interact with the background
magnetic waves and/or turbulence, i.e., the so-called ``injection
problem'' must be considered \citep[see, for
example,][]{JE91,Malkov98}.
The \mc\ techniques we describe here make the simple assumption that
all particles, regardless of energy, interact in the same way, i.e.,
all particles scatter elastically and isotropically in the local
plasma frame with a mean free path proportional to their gyroradius.
These techniques and assumptions have been used to calculate nonlinear
effects in \nonrel\ collisionless shocks for a number of years with
good success comparing model results to spacecraft observations
\citep[e.g.,][]{EE84,EMP90,EJB99}.

Early work on relativistic shocks was mostly analytical in the test
particle approximation \citep[e.g.,][]{BM76,Peacock81,KS87a,HD88},
although the analytical work of \citet{SK87} explored modified shocks.
Test-particle Monte Carlo techniques for \rel\ shocks were developed
by \citet{KS87b} and \citet{EJR90} for parallel, steady-state shocks,
i.e., those where the shock normal is parallel to the upstream
magnetic field, and extended to include oblique magnetic fields by
\citet{Ostrow91}.  Some preliminary work on modified relativistic
shocks using Monte Carlo techniques was done by
\citet{EllisonJapan91,EllisonPoland91}.

The most important results from the theory of test-particle
acceleration in \ultrarel\ shocks are:
(i) regardless of the state of the unshocked plasma, some particles
can pick up large amounts of energy 
$\Delta E \sim \gamsk^2mc^2$
in their first shock crossing cycle \citep{Vietri95}, but will receive
much smaller energy boosts ($\ave{E_f/E_i} \sim 2$) for subsequent
crossing cycles \citep[e.g.,][]{GA99,AGKG2001}\footnote{$E_i$ ($E_f$)
is the particle energy at the start (end) of an upstream to downstream
to upstream (or a downstream to upstream to downstream) shock crossing
cycle.};
(ii) the shock compression ratio, defined as $r \equiv u_0 / u_2$,
tends to $3$ as $u_0 \rightarrow c$ \citep[e.g.,][]{Peacock81,Kirk88},
where $u_2$ is the flow speed of the shocked plasma measured in the
shock frame;\footnote{Note that the density ratio across the relativistic
shock
$\rho_2/\rho_0 \ne u_0/u_2$, in contrast with
\nonrel\ shocks, because
the Lorentz factors associated with the relativistic flows modify the
particle flux jump condition. Here and elsewhere we use the subscript 0 (2)
to indicate far upstream (downstream) values.}
 and
(iii) a so-called `universal' spectral index, $\sigma \sim 4.2-4.3$
(in Eq.~\ref{eq:TPpowerlaw}) exists in the limits of $\gamsk \gg
1$ and $\deltheta \ll 1$, where $\deltheta$ is the change in direction
a particle's momentum vector makes at each pitch angle scattering
\citep[e.g.,][]{BedOstrow98,AGKG2001}.

We find that these results are modified in mildly \rel\ shocks, even
in the test-particle approximation, and in fully \rel\ shocks (at
least for 
$\gamma_0 \lesssim 10$) when the backreaction of the
accelerated particles is treated self-consistently, which causes the
shock to smooth and the compression ratio to change from test-particle
values.
In mildly \rel\ shocks, $f(p)$ remains a power law in the \TP\
approximation but both $r$ and $\sigma$ depend on the shock Lorentz
factor, $\gamsk$.  When efficient particle acceleration occurs in
mildly \rel\ shocks 
(i.e., $\gamsk \lesssim 3$), large increases in $r$ can
result and a power law is no longer a good approximation to the
spectral shape.  In these cases, we determine the compression ratio by
balancing the momentum and energy fluxes across the shock with the
\mc\ simulation.
For larger Lorentz factors, accelerated particles smooth the
shock structure just as they do in slower shocks,
but $r$ approaches 3 as $\gamsk$ increases.
In general, efficient particle acceleration results in spectra very
different from the so-called `universal' power law found in the
test-particle approximation unless $\gamsk \gtrsim 10$.

\section{\mc\ Model}

The techniques we use are essentially identical to those described in
\citet{EBJ96} and \citet{EJB99}. The differences are that the code has
been made fully \rel\ and only results for parallel shocks with
pitch-angle diffusion are presented here.
The code is steady-state, includes a uniform magnetic
field, and moves particles in helical orbits.  We assume the \alf\
Mach number is large, i.e., we neglect any effects from \alf\ wave
heating in the upstream precursor. 
This also means we neglect the second-order acceleration of particles
scattering between oppositely propagating \alf\ waves. 
Such an effect in \rel\ plasmas with strong magnetic fields is
proposed for nonlinear
particle acceleration in GRBs by \citet{Pelletier99}
\citep[see also][]{PellMar98}.

The pitch angle diffusion is
performed as described in \citep{EJR90}.  That is, after a small
increment of time, $\deltime$, a particles' momentum vector,
$\mathbf{p}$, undergoes a small change in direction, $\deltheta$.  If
the particle originally had a pitch angle, $\theta$ (measured relative
to the shock normal direction), it will have a new pitch angle
$\theta'$ such that
\begin{equation}
\label{eq:cosprime}
\cos{\theta'} = \cos{\theta} \, \cos{\deltheta} +
\sqrt{1 - \cos^2{\theta}} \sin{\deltheta} \cos{\phi}
\ ,
\end{equation} 
where $\phi$ is the azimuth angle measured with respect to the
original momentum direction. All angles are measured in the local
plasma frame. If $\deltheta$ is chosen randomly from a uniform
distribution between 0 and $\delmax$ and $\phi$ is chosen from a
uniform distribution between 0 and $2 \pi$, the tip of the momentum
vector will perform a random walk on a sphere of radius $p$.  As shown
by \citet{EJR90}, the
angle $\delmax$ is determined by
\begin{equation}
\label{eq:thetamax}
\delmax =
\left ( 6 \, \deltime/t_c \right )^{1/2} =
\left ( 12 \pi/ N \right )^{1/2}
\ ,
\end{equation}
where $N = \tau_g / \deltime \gg 1$ is the number of gyro-segments,
$\deltime$, dividing a gyro-period $\tau_g = 2 \pi r_g/v$. The time
$t_c$ is a ``turn around'' time defined as $t_c = \lambda / v$, where
$\lambda$ is the particle mean free path.  The mean free path is taken
to be proportional to the gyroradius $r_g = pc/(QeB)$ ($e$ is the
electronic charge, $Q$ is the ionic charge number, and $B$ is the
local uniform magnetic field), i.e.,
$\lambda = \eta \, r_g$,
where $\eta$ determines the strength of scattering. In all of the
examples given here we set $\eta=1$, or, in other words, the strong
scattering Bohm limit is assumed.

For a downstream particle to return upstream, it's velocity vector
must be directed within a cone with opening angle $\theta_2$ such that
$|v_2\cos{\theta_2}| > u_2$, where $v_2$ and $\theta_2$ are measured
in the downstream frame and $\theta_2=0^{\circ}$ is in the
$-x$-direction, i.e., along the shock normal direction.  
For 
fully relativistic
shocks with $v_2 \simeq c $ and $u_2
\simeq c/3$, $\cos{\theta_2} \gtrsim 1/3$ for a downstream particle to
cross the shock into the upstream region.
When the particle enters the upstream region it 
must satisfy essentially the same constraint, i.e.,
$|v_0 \cos{\theta_0}| > u_0$, 
where now $v_0$ and $\theta_0$ are measured in the upstream
frame.\footnote{In the \TP\ approximation, $u_0$ is just the shock
speed. In nonlinear shocks, the flow speed just upstream from the
subshock at $x=0$ will be less than the far upstream shock speed,
$u_0$, as measured in the shock reference frame.}
Since both the particle and shock
have high Lorentz factors, we can write
\begin{equation}
\cos{\theta_0} = u_0/v_0 = 
\frac{ \left ( 1 - 1/\gamma^2_0 \right )^{1/2}}
{\left (1 - 1/ \gamma^2_v \right )^{1/2}} \simeq
\frac{1 - 1/(2 \gamma^2_0)}{1- 1/(2\gamma^2_v)}
\ ,
\end{equation}
where $\gamma_v \equiv [1-(v_0/c)^2]^{-1/2}$ is the particle Lorentz
factor. Since $\cos{\theta_0}
\simeq 1 - \theta_0^2/2$ for small $\theta_0$, we have
\begin{equation}
\theta_0^2 \simeq \frac{1}{\gamma^2_0} - \frac{1}{\gamma^2_v}
\ .
\end{equation}
For ultrarelativistic particles with $\gamma_v \gg \gamma_0$, 
$\theta_0 \simeq 1/\gamma_0$ \citep[e.g.,][]{GA99},
but $\theta_0$ can be much smaller for mildly \rel\ particles.

\begin{figure}[!hbtp]              
\epsscale{0.55} 
\plotone{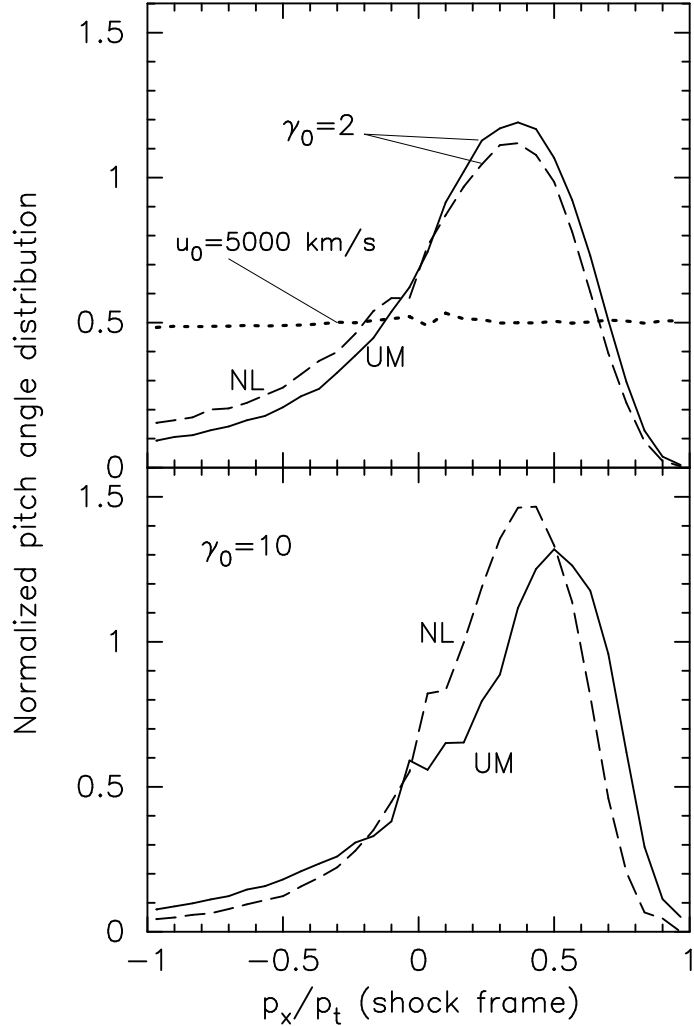} 
\figcaption{Distribution of the cosine of the pitch angle, i.e.,
$p_x/p_t$, for particles crossing $x=0$. The solid and dashed curves
in the top panel are for a shock with Lorentz factor $\gamma_0=2$. The
dotted curve in the top panel is for a \nonrel\ shock with speed
$u_0=5000$ \kmps. The bottom panel shows curves for a fully \rel\
shock with $\gamma_0=10$. In all cases, the curves are normalized so
that the areas under them is 1 and the $x$-component of momentum,
$p_x$, is positive when directed downstream. The nonlinear (NL) and
unmodified (UM) shock results are labeled.
\label{fig:angle}
}
\end{figure}

In order to re-cross into the downstream region, particles must
scatter out of the upstream cone defined by $\theta_0$ and
\citet{AGKG2001} show that most particles are only able to change the
angle they make with the upstream directed shock normal by
$|\deltheta| \sim \theta_0$ before being sweep back downstream, making
the distribution of shock crossing particles highly anisotropic.
Therefore, if the shock Lorentz factor $\gamsk \gg 1$, a larger
fraction of particles re-cross the shock into the downstream region
with highly oblique angles (as measured in the shock
frame) compared to lower speed shocks (see Fig.~\ref{fig:angle}
discussed below).
Particles crossing at such oblique angles receive smaller energy gains
than would be the case for an isotropic pitch angle distribution and
\citet{AGKG2001} go on to show that $\ave{E_f/E_i} \sim 2$ for a shock
crossing cycle (after the first one).

\begin{figure}[!hbtp]              
\epsscale{0.55} \plotone{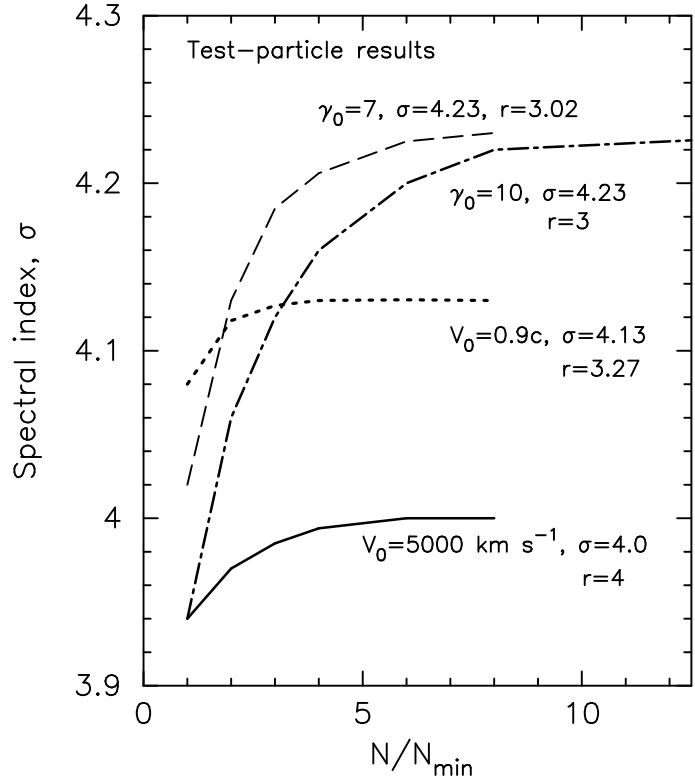} 
\figcaption{Power law spectral index, $\sigma$, versus number of
gyro-segments, $N$, for various test-particle shocks as labeled. In
all cases, as $N$ is increased the spectral index converges and we
obtain the known results of $\sigma \simeq 4$ for \nonrel\ shocks
with $r=4$, and $\sigma \simeq 4.23$ for fully \rel\ shocks with
$r=3$. The parameter $\Nmin$ is defined in Eq.~(\ref{eq:gyroeq}).
\label{fig:gyro}
}
\end{figure}

Considering these constraints, we require $\delmax < \theta_0$, 
or
\begin{equation}
\label{eq:gyroeq}
N > \Nmin = 12\pi\gamsk^2
\ ,
\end{equation}
and find (as shown in Fig.~\ref{fig:gyro}) that the power law
spectral index, $\sigma$, asymptotically approaches a maximum value as
$N$ is increased.  If $N$ is less than the value required for
convergence (and the gyro-segments are too large), the distribution
will be flatter than produced with the convergent value of $N$ because
more particles are able to cross from upstream to downstream with
$\thetaSK \ll 90^\circ$ and receive unrealistically large energy
boosts.
This effect has long been known from the comparison of pitch-angle
diffusion to large-angle scattering in \rel\ shocks
\citep[e.g.,][]{KS87b, EJR90}.
For all of the examples reported on here, $N$ is chosen large enough
so it makes no difference if $\deltheta$ is chosen uniformly between
$0^\circ$ and $\delmax$ or if $\cos{\deltheta}$ is chosen uniformly
between $\cos{\delmax}$ and 1.
Fig.~\ref{fig:gyro} shows how our results depend on $N$ for shock
speeds ranging from fully \nonrel\ to fully \rel.  In all cases, as
$N$ is increased the spectral index approaches a maximum and for
$\gamsk \gtrsim 7$ we obtain the well known result $\sigma =
4.2$--$4.3$.
The fact that the computation time for the Monte Carlo simulation scales
as $N$ and $N \propto \gamsk^2$ places limits on modeling \ultrarel\
shocks with this technique.

In Fig.~\ref{fig:angle} we compare pitch-angle distributions
(measured in the shock reference frame) of particles crossing the
shock.
The curves are normalized such that the area under each curve equals
one and the $x$-component of particle momentum in the shock frame,
$p_x$, is positive when directed downstream to the right (see
Fig.~\ref{fig:profG2} for the shock geometry). Here, $p_t$ is the
magnitude of the total particle momentum also measured in the shock
frame.
In the top panel, we compare an unmodified (UM) and nonlinear (NL)
mildly \rel\ shock ($\gamsk=2$) with a \nonrel\ one ($u_0=5000$
\kmps).
Particles crossing the $\gamsk=2$
shock are highly anisotropic with $p_x/p_t$ strongly peaked near $\sim
0.35$. In the \nonrel\ shock, the particles are nearly isotropic
except for a slight flux-weighting effect. 
There is little difference
in the distributions between the UM and NL shocks.
In the bottom panel, we show the pitch-angle distributions for UM and
NL shocks with $\gamsk=10$. While the distributions are somewhat more
sharply peaked, they are quite similar to those for $\gamsk=2$ and
show relatively small variations between the UM and NL shocks.

The main difference between our present code and an earlier code used
by \citet{EJR90} to model test-particle \rel\ shocks is that the
previous code used a guiding center approximation with an emphasis on
large-angle scattering rather than the more explicit orbit calculation
of pitch-angle diffusion used here.  Other than the far greater range
in $\gamsk$ and the nonlinear results we now present, the work of
\citet{EJR90} is consistent with the work presented here.

The particle transport is performed as follows.  Particles of some
momentum, $\ppf$, (measured in the local plasma frame) are injected
far upstream from the shock and pitch-angle diffuse and convect until
they cross a grid zone boundary, i.e., a dividing plane in the
simulation between regions with different bulk flow speeds
\citep[see][for a full discussion]{EBJ96}. For unmodified shocks there
is only one grid zone boundary which divides the upstream and
downstream regions, but for nonlinear shocks the bulk flow speed
changes in small steps, each separated by a boundary, from $u_0$ far
upstream to $u_2$ downstream.
In the unmodified case, the shock thickness is essentially zero (i.e.,
shorter than the distance a particle diffuses in $\deltime$) but in
the nonlinear, modified case, the shock precursor extends over the
entire region of varying bulk flow speeds and a small scale
``subshock'' (at position $x=0$ in our simulation) exists where most
of the entropy production occurs.
When a particle crosses a grid zone boundary, $\ppf$ is transformed to
the new local frame moving with a new speed relative to the subshock
and the particle continues to scatter and convect.  Each particle is
followed until it leaves the system in one of three ways. It can
convect far downstream and not return to the subshock, it can obtain a
momentum greater than some $\pmax$ and be removed, or it can diffuse
far enough upstream to cross an upstream free escape boundary (FEB)
and be removed. Both $\pmax$ and the position of the FEB are free
parameters in our model \citep[see][for a discussion of the
self-consistent determination of the maximum
particle energy in \nonrel\ supernova remnant shocks]{BerezV97}.
For our nonlinear calculations, we iterate the shock structure and
compression ratio until the number, momentum, and energy fluxes are
conserved across the shock.  This procedure has been detailed many
times for \nonrel\ shocks \citep[see][and references therein]{EBJ96}
and the modifications required for \rel\ shocks were given in
\citet{ER91}.

To avoid excessive computation, we use a probability of return
calculation as described in detail in \citet{EJR90}.  That is, we use
the standard expression obtained by \citet{Peacock81},
\begin{equation}
\label{eq:ProbRet}
P_R = \left ( \frac{v - u_2}{v + u_2} \right )^2
\ ,
\end{equation}
to determine the probability, $P_R$, that a particle, having crossed a
particular point in the uniform downstream flow, will return back
across that point.
Eq.~(\ref{eq:ProbRet}) is fully relativistic and independent of
the diffusive properties of the particles as long as they are
isotropic in the $u_2$ frame. We ensure this isotropy by only
applying Eq.~(\ref{eq:ProbRet}) once a particle has diffused
several mean free paths downstream from the shock.
For \ultrarel\ shocks with $v\simeq \Vsk \simeq c$ and $r \simeq 3$,
Eq.~(\ref{eq:ProbRet}) gives $P_R \simeq 0.25$.  

For clarity, we note that this is not the same probability, $\Pret$,
that is used by \citet{AGKG2001} to determine the \TP\ power law
index, i.e.,
\begin{equation}
s = 1 + \frac{\ln (1/\Pret)}{\ln \ave{E_f/E_i}}
\ ,
\end{equation}
where $N(E) \propto E^{-s}$ and, for fully \rel\ particles, $s =
\sigma - 2$. The quantity $\ave{E_f/E_i}$ is the average energy ratio
for a particle undergoing a shock crossing cycle.
Although not explicitly stated in \citet{AGKG2001}, it is clear from
the context that $\Pret$ is calculated just behind the shock where the
particle distribution is highly anisotropic.  In this case, particles
that have just crossed from upstream to downstream will be more likely
to recross back into the upstream region than indicated by
Eq.~(\ref{eq:ProbRet}) because their pitch angles are more likely
to be highly oblique relative to the shock normal than in the
isotropic distributions further downstream (compare the solid or
dashed curves to the dotted curve in the top panel of
Fig.~\ref{fig:angle} discussed below). For $\gamsk=10$,
\citet{AGKG2001} find $\Pret = 0.435 \pm 0.005$ and $\ave{E_f/E_i} =
1.97 \pm 0.01$ giving the standard result $s = 2.230 \pm 0.012$.  As
shown in Fig.~\ref{fig:gyro}, our unmodified results are consistent
with this spectral index for $\gamsk \gtrsim 7$.

\begin{figure}[!hbtp]              
\epsscale{0.55} \plotone{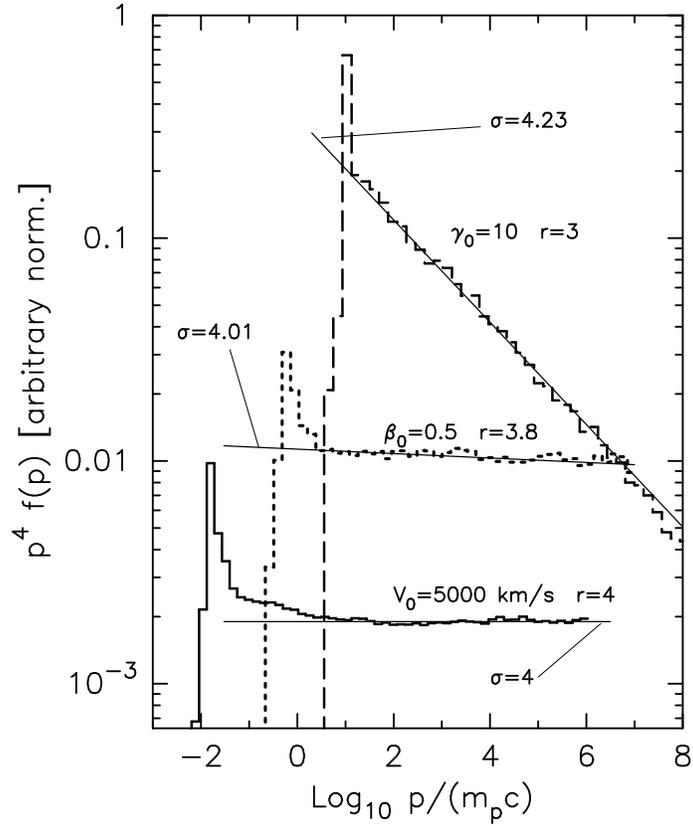} 
\figcaption{Particle spectra, $p^4 f(p)$, versus momentum for various
unmodified shocks with speeds as indicated. The test-particle
compression ratios, $r$, and spectral indices, $\sigma$, are
noted. The far upstream plasmas in the 5000 \kmps, and 
$\beta_0=u_0/c=0.5$
shocks are thermal at $10^6$ K. The $\gamma_0=10$ shock has a far
upstream plasma which is a $\delta$-function at 1 MeV.  All spectra
are calculated at the shock, in the shock frame, and the relative
normalization is arbitrary.
\label{fig:specTP}
}
\end{figure}

\section{Test-Particle Results}

In Fig.~\ref{fig:specTP} we show particle distributions for
unmodified shocks with speeds ranging from fully \nonrel\ ($\Vsk =
5000$ \kmps) to mildly \rel\ ($\beta_0 = u_0/c = 0.5$) to fully \rel\
($\gamsk = 10$). The \nonrel\ distribution matches the standard \TP\
Fermi result of $\sigma = 4$ for $r=4$ and the fully \rel\ result is
consistent with the well-known limit of $\sigma \too 4.2 - 4.3$ as
$\gamsk \too \infty$ for $r=3$. In the trans-relativistic regime, both
the compression ratio and the spectral index vary with $\Vsk$.
For the $\Vsk = 0.5c$
distribution shown in Fig.~\ref{fig:specTP}, we have determined the
compression ratio by balancing the mass, momentum, and energy fluxes
across the shock in the test-particle limit, i.e., by ignoring any
effects from the accelerated particles. This technique is described in
detail in \citet{ER91}.  We find $r = 3.8 \pm 0.1$ and the spectral
index is $\sigma \simeq 4.01$, slightly flatter than $3 \, r / (r-1)
\simeq 4.07$, the \nonrel\ result.

\begin{figure}[!hbtp]              
\epsscale{0.55}
\plotone{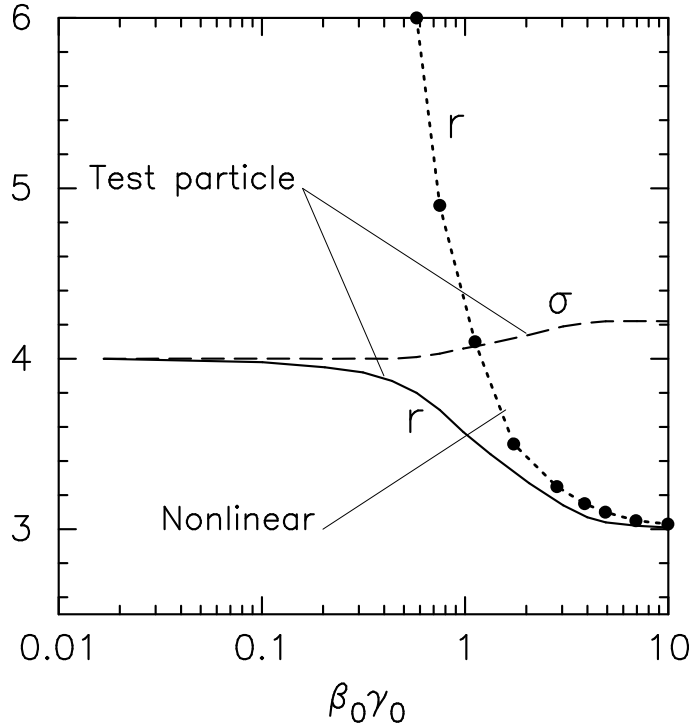}
\figcaption{The solid line is the compression ratio, $r$, and the
dashed line is the spectral index, $\sigma$, for unmodified (i.e.,
\TP) shocks. The solid dots show $r$ for shocks undergoing efficient
particle acceleration. In all cases, $r$ is determined for each
$\beta_0 \gamsk$ by balancing the momentum and energy fluxes across
the shock. The maximum cutoff momentum, $\pmax$, is unimportant for
the \TP\ shocks, but is set to $10^4\, \beta_0 \gamsk \, m_p c$ for
the nonlinear shocks.
\label{fig:ratio}
}
\end{figure}

Fig.~\ref{fig:ratio} shows the compression ratio as
a function of $\beta_0 \gamsk$ (solid curve), still ignoring the
effects of accelerated particles.  
The compression ratio is determined self-consistently by balancing the
momentum and energy fluxes across the shock with no restriction on the
shocked or unshocked adiabatic index.
As expected, $r$ decreases smoothly
from $4$ for fully \nonrel\ shocks to $\sim 3$ for fully \rel\ shocks.
The power law index, $\sigma$, is also shown (dashed curve)
and this varies slowly from $\sigma =4$ to $\sigma \simeq 4.23$
between the two extremes.
For comparison, we show $r$ (solid dots) for shocks undergoing
efficient particle acceleration. For these points, all shock
parameters 
except $\beta_0 \gamsk$ and $\pmax$ are kept constant. For the
nonlinear shocks, the maximum cutoff momentum is set to $\pmax= 10^4
\, \beta_0 \gamsk \, m_p c$ in all cases.\footnote{The value of
$\pmax$ can have a large influence on the shock characteristics at low
$\beta_0 \gamsk$ because $r$ is large enough (and $\sigma$ small
enough)
that particles escaping
at $\pmax$ are dynamically important. When $\beta_0 \gamsk$ becomes
large enough so that $r$ drops below $\sim 4$, $\pmax$ is no longer an
important parameter for the shock structure.}
We discuss the nonlinear results in detail below, but here
only emphasize that $r$ in nonlinear shocks will be larger than the
\TP\ value for $\gamsk \lesssim 10$.
The test-particle results shown in Fig.~\ref{fig:ratio} are in close
agreement with recent analytic results of 
\citet{KGGA2000} and \citet{Gallant2002}.
To our knowledge, no analytic results exist for nonlinear, \rel\ shocks.

\begin{figure}[!hbtp]              
\epsscale{0.55}
\plotone{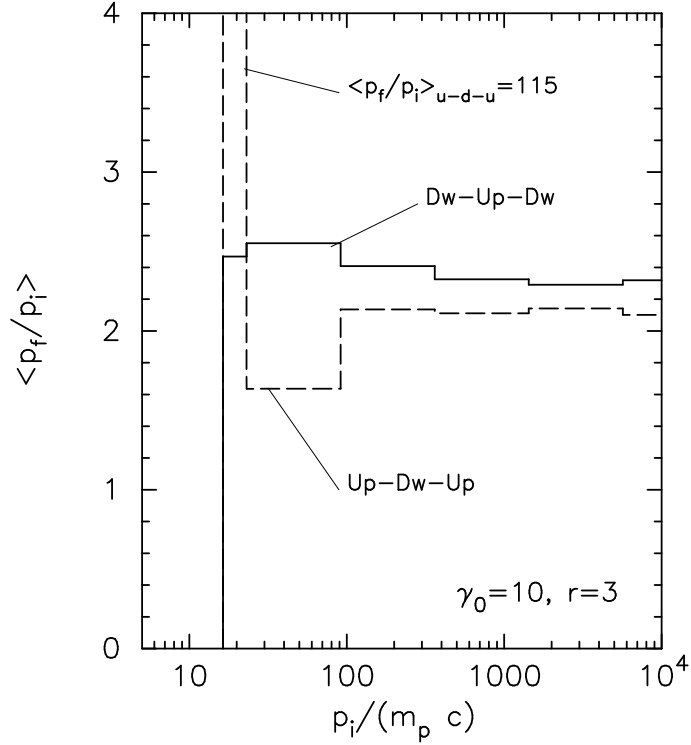}
\figcaption{Average ratios of final (f) to initial (i) momentum for
downstream to upstream to downstream and upstream to downstream to
upstream shock crossing cycles. The histograms are for a test-particle
shock with $\gamma_0=10$ and $r=3$. Note the large momentum gain
($\UDUave = 115$) in the first shock crossing cycle.
Note also that this particular value depends on our choice of the 
injection momentum.
\label{fig:pfpi}
}
\end{figure}

In Fig.~\ref{fig:pfpi} we show the average ratios of momenta
(measured in the local plasma frame) for particles executing upstream
to downstream to upstream cycles across the shock, \hbox{$\UDUave$},
and downstream to upstream to downstream cycles, \hbox{$\DUDave$}.
These results are for $\gamsk=10$ ($r=3$) and show a slight momentum
dependence at low momenta but converge to $\ave{p_f/p_i} = 2.2 \pm
0.1$ at high momenta. This value is close to $\ave{E_f/E_i} = 1.97 \pm
0.01$ reported by \citet{AGKG2001} for $\gamsk=10$.
As mentioned above, in the first upstream to downstream to upstream
cycle, particles achieve a large boost in momentum as indicated in the
figure.

The difference between our value of $\ave{p_f/p_i}$ at large $p_i$ and
that of \citet{AGKG2001} is greater than the uncertainties and
probably stems from the different assumptions made in the simulations.
As discussed above, $\ave{p_f/p_i}$ depends critically on the average
angle a particle makes when crossing the shock. While the large
majority of particles in our simulations gain energy by crossing from
downstream to upstream and then immediately (within a few
$\deltime$'s) re-crossing back into the downstream region at oblique
angles, a few manage to diffuse farther upstream (see
Fig.~\ref{fig:traj09c} below). When these particles re-cross the
shock into the downstream region, they can do so at flatter angles and
receive larger energy gains. We speculate that differences in how
these few particles are treated in the simulations might produce the
differences in $\ave{p_f/p_i}$.

\begin{figure}[!hbtp]              
\epsscale{0.55}
\plotone{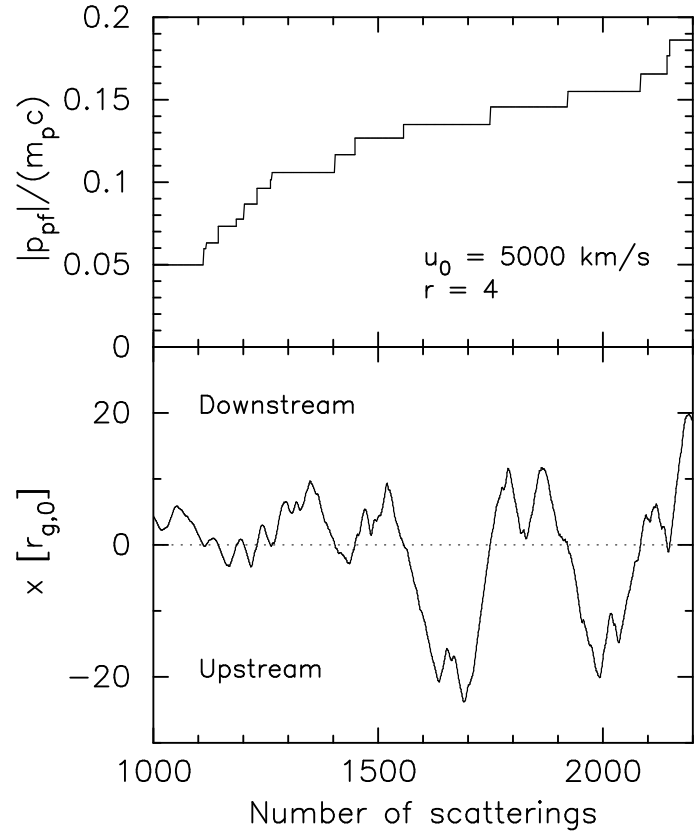}
\figcaption{Particle trajectory (lower panel) and momentum (upper
panel) in an unmodified shock of speed $u_0=5000$ \kmps. The momentum
is calculated in the local plasma frame, either upstream or downstream
from the shock.
\label{fig:traj5000}
}
\end{figure}

As an illustration of how particles interact with nonrelativistic and
relativistic unmodified shocks, we shown trajectories for two
individual particles in Figs.~\ref{fig:traj5000} and
\ref{fig:traj09c}. The lower panel in each figure shows a trace of the
particle trajectory and the upper panels show the particle momentum,
always measured in the local plasma frame. For the \nonrel\ shock
(Fig.~\ref{fig:traj5000}), the speed of the particle is far greater
than the shock speed and it diffuses easily on both sides of the
shock. When it crosses $x=0$, it does so nearly isotropically (except
flux weighting makes crossings with flat trajectories slightly more
likely) and essentially always gains momentum. The momentum gain in a
single shock crossing is small, but a particle can stay in the system
for many crossings.

When the shock speed, $u_0$, is close to $c$, the particle will be
convected downstream much more rapidly than in \nonrel\ shocks and few
particles will be able to cross the shock many times. However,
downstream particles that do manage to cross the shock into the
upstream region do so with much flatter trajectories, as discussed
above, and can receive large momentum boosts in a single shock
crossing due to the shock's Lorentz factor (note the logarithmic scale
in the top panel of Fig.~\ref{fig:traj09c}). In a typical shock
crossing cycle, downstream particles gain momentum when they cross
into the upstream region (see positions labeled $a$ and $b$ in
Fig.~\ref{fig:traj09c}), lose momentum when they cross back
downstream because they cross with oblique pitch angles,
but end up with a net momentum boost.
However, as shown by the position labeled $c$, it is possible for a
particle to diffuse farther upstream before being convected back to
the shock.  In this case, it can cross the shock with a flat
trajectory and gain momentum upon entering the downstream region.
If the acceleration is efficient, the few particles that diffuse far
upstream carry enough pressure to produce the shock smoothing we
discuss next.

\begin{figure}[!hbtp]              
\epsscale{0.55}
\plotone{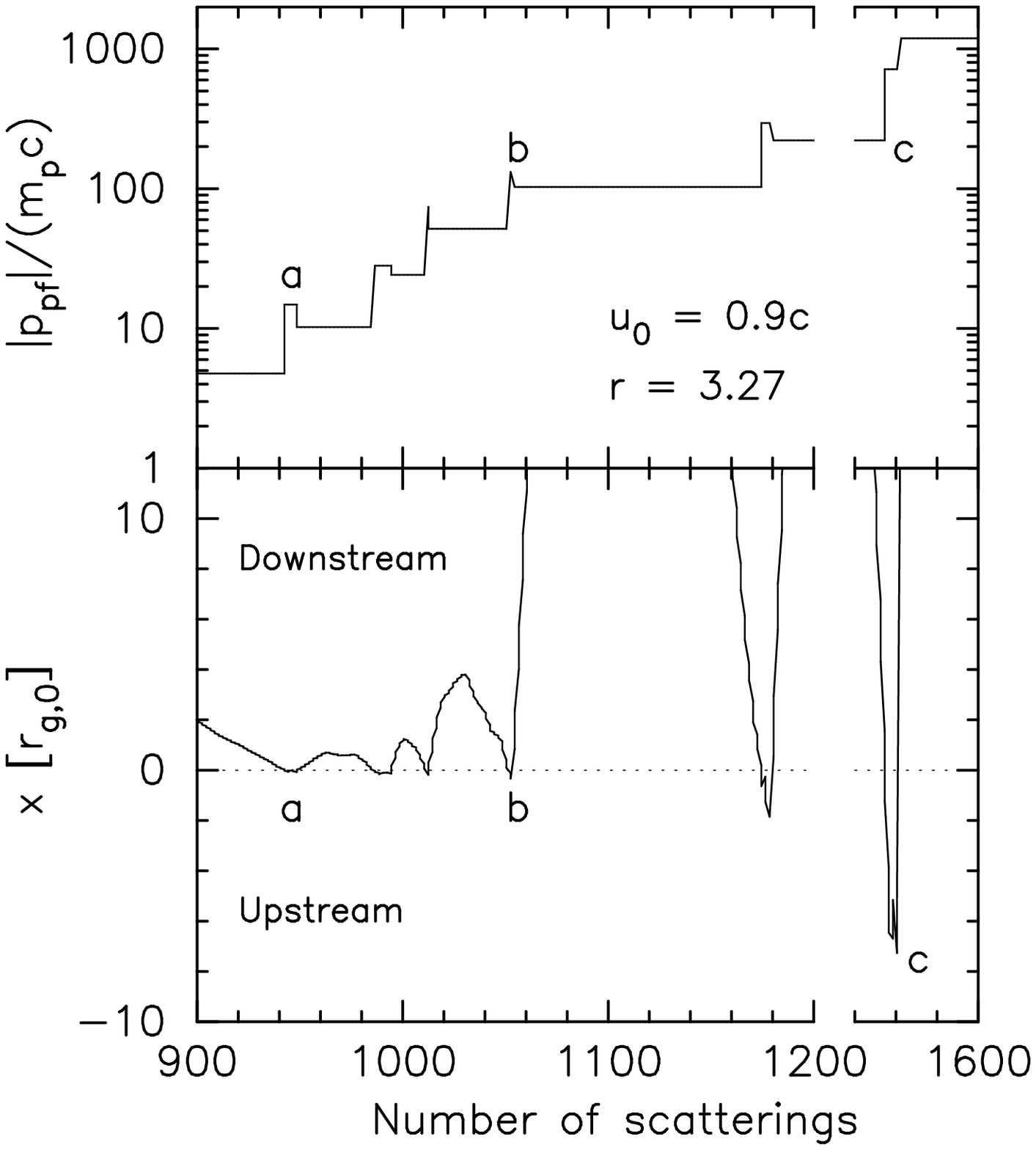}
\figcaption{Trajectory and momentum for a particle in an unmodified
shock with speed $u_0=0.9c$. Note that the horizontal axis is split at
1200 scatterings.
\label{fig:traj09c}
}
\end{figure}

\section{Non-Linear Results}

As noted by \citet{AGKG2001}, the large energy boost particles receive
in their initial crossing of the shock provides a natural injection
process for further acceleration and suggests that \rel\ shocks may be
efficient accelerators.
However, just as with \nonrel\ shocks, efficient acceleration limits
the use of test-particle approximations and requires that the
nonlinear backreaction of the accelerated particles be treated
self-consistently \citep[e.g.,][]{JE91}.
These nonlinear effects will result in a smoothing of the shock and a
change in the overall shock compression ratio, just as they do in
\nonrel\ shocks.
The differences between test-particle and nonlinear results, for 
the parameter ranges we have investigated, are large enough to
produce spectra noticeably different from the often quoted $N(E)
\propto E^{-2.3}$ and to influence applications to GRB models where
high particle acceleration 
efficiencies are assumed.  We illustrate this with two
examples, one mildly \rel\ ($\gamsk = 1.4$) and one more fully \rel\
($\gamsk = 10$).  The far upstream conditions have relatively little
influence on our results as long as $\gamma_p \ll \gamma_0$, where
$\gamma_p$ is the plasma frame Lorentz factor of the far upstream,
injected particles. For concreteness, in both examples we take the far
upstream plasma to be a thermal distribution of protons
at a temperature of $10^6$ K.

\subsection{Mildly relativistic shock: $\gamma_0 = 1.4$}

\begin{figure}[!hbtp]              
\epsscale{0.55}
\plotone{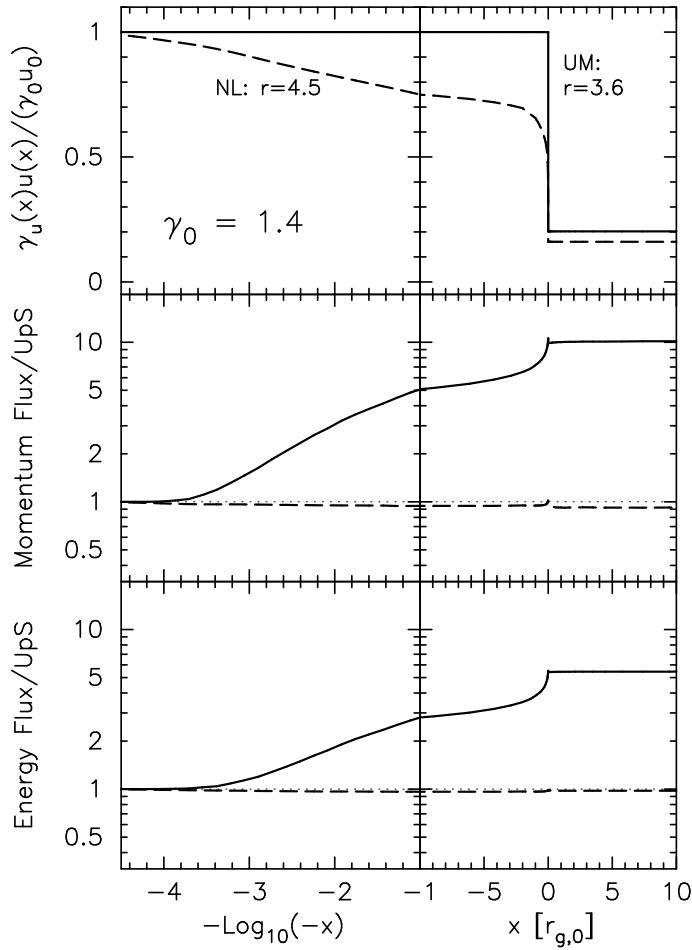}
\figcaption{Unmodified (UM) and nonlinear (NL) shock profiles, i.e.,
$\gamux u(x)$, and momentum and energy fluxes versus position,
$x$. All quantities are scaled to far upstream values and in all
panels the solid curves are results from unmodified shocks and the
dashed curves are nonlinear results. The NL momentum and energy fluxes
are 3 to 4\% below the far upstream values because particles escape at
a maximum momentum, 
$\pmax = 10^4 \beta_0 \gamsk m_p c \simeq
10^{13}$ eV/c. We have used $N/\Nmin \sim 10$ in both cases.
\label{fig:profG2}
}
\end{figure}

Fig.~\ref{fig:profG2} shows unmodified and nonlinear shock
structures for $\gamsk = 1.4$. The top panel shows $\gamux \, u(x)$,
where $\gamux = \{1 - [u(x)/c]^2\}^{-1/2}$, the middle panel is
momentum flux, and the bottom panel is the energy flux, all scaled to
far upstream values.  All curves are plotted versus $x$, where $x=0$
is the position of the sharp subshock.  A logarithmic scale is used
for $x < -10 \, \rgz$ and a linear scale is used for $x > -10\,\rgz$,
where $\rgz \equiv m_p u_0/(eB)$.\footnote{In these parallel
shock simulations where we ignore \alf\ wave production, the magnetic
field, $B$, has no other effect than setting arbitrary length and time
scales.}
In each panel, the solid curve is from an unmodified shock with $r
\simeq 3.6$ (see Fig.~\ref{fig:ratio}), while the dashed curve is
the momentum and energy flux conserving result.

For our pitch angle diffusion model, where particles interact
elastically and isotropically in the local frame according to
Eq.~(\ref{eq:cosprime}) and the discussion following it,
particles are accelerated efficiently enough at the unmodified shock
that the momentum and energy fluxes are not conserved and rise well
above the allowed far upstream values. In order to conserve these
fluxes, the shock structure must be smoothed and the overall
compression ratio increased above the test-particle value.  Our
computational scheme calculates this compression ratio and flux
conserving profile and the result is the dashed curve in the top panel
with the corresponding momentum and energy fluxes in the middle and
bottom panels.

The source of the non-conservation of momentum and energy is the
efficient acceleration of particles by the sharp flow speed
discontinuity. While the actual injection and acceleration efficiency
depends on our particular pitch angle diffusion model, once our
scattering assumptions are made, the kinematics determine the
injection and acceleration of the particles without additional
parameters.\footnote{Other input parameters, such as the Mach number
and $\pmax$ or the position of the FEB, influence the acceleration
efficiency, but these are ``environmental'' parameters rather than
parameters needed to describe the plasma interactions.}
Of course it would have been possible to make assumptions which
resulted in an acceleration efficiency low enough that momentum and
energy are approximately conserved without significantly smoothing the
shock structure or changing the compression ratio from the
test-particle value. For example, we could have only allowed shocked
particles above some Lorentz factor $\gaminj$ to recross into the
upstream region or arbitrarily restricted the number of particles that
recrossed into the upstream to a small fraction, $\fracinj$, of all
downstream particles. By making $\fracinj$ low enough or $\gaminj$
high enough we could make the efficiency as low as we wanted.
However, there are at least three reasons for not making such an
assumption. The first is that restricting the acceleration efficiency
requires additional parameters (i.e., $\gaminj$ and/or $\fracinj$) to
those needed to describe pitch angle
diffusion.\footnote{Alternatively, a far more complex model of the
plasma interactions can be postulated than done here, inevitably
requiring additional parameters \citep[e.g.,][]{Malkov98}.}
The second is that models with {\it inefficient} acceleration will not
help explain GRBs (or other objects) that require high efficiencies.
If \rel\ shocks are inefficient accelerators they are not very
interesting. If they are efficient, they will have a qualitative
resemblance to the results we show even if, as is likely, the
actual plasma processes are far more complex than the simple model we
use.
A third, perhaps less compelling, reason is that identical
scattering assumptions as used here have been used for some time in
\nonrel\ shocks and shown to match both spacecraft observations
\citep[e.g.,][]{EMP90} and hybrid plasma simulations
\citep[e.g.,][]{EGBS93} of collisionless shocks.

The increase in compression ratio from $r \simeq 3.6$ to $\simeq 4.5$
shown in Fig.~\ref{fig:profG2} comes about, in part, because the
particles escaping at $\pmax$ carry away momentum and energy fluxes
which make the shocked plasma more compressible.
This effect is countered to some degree by the fact that the
self-consistent shock produces a downstream distribution with a
smaller fraction of \rel\ particles than the unmodified shock so that
the downstream adiabatic index
$\Gamma > 4/3$. This tends to
produce a smaller compression ratio.
The escaping fluxes show up as a lowering of the dashed curves below
the far upstream values, as shown in the bottom two panels of
Fig.~\ref{fig:profG2}, and amount to about 3\% of the far upstream
values for both momentum and energy.  Including the escaping fluxes,
the momentum and energy fluxes are conserved to within a few percent
of the far upstream values.
While the changes seen here are similar to those seen and discussed
for many years in efficient, \nonrel\ shock acceleration
\citep[see][and references therein]{BE99}, it must be noted that there
are no known analytic expressions relating escaping fluxes and
$\Gamma$ to $r$ in this \transrel\ regime. One obvious difference is
that for \nonrel\ shocks, the escaping momentum flux is generally much
less than the escaping energy flux (when both are measured as
fractions of incoming flux) since $\rho_e v_e^3/(\rho_0 u_0^3) \gg
\rho_e v_e^2/(\rho_0 u_0^2)$ when $v_e \gg u_0$
\citep[see][]{Ellison85}. Here, $v_e$ is the velocity of the escaping
particle.  In \rel\ shocks, $v_e \sim u_0 \sim c$ so the escaping
fluxes are about equal as shown in Fig.~\ref{fig:profG2}.

\begin{figure}[!hbtp]              
\epsscale{0.55}
\plotone{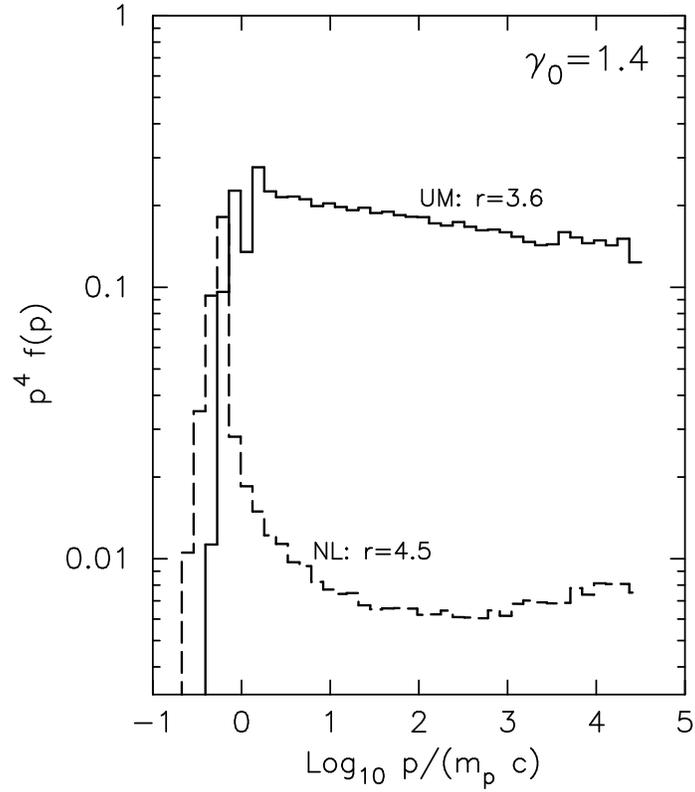}
\figcaption{Particle distributions, $p^4 f(p)$, for the shocks
shown in Fig.~\ref{fig:profG2}. The nonlinear spectrum (dashed
curve) shows the distinctive concave shape seen in efficient
\nonrel\ shock acceleration, and has a greater fraction of low
momentum particles than the spectrum from the unmodified shock.
As in Fig.~\ref{fig:specTP}, the spectra are calculated at the shock
in the shock frame and truncated with a $\pmax$. Unlike
Fig.~\ref{fig:specTP}, the normalization here shows the actual
acceleration efficiency.
\label{fig:specG2}
}
\end{figure}

In Fig.~\ref{fig:specG2} we plot $p^4 f(p)$ for our $\gamma_0=1.4$
shock. The solid curve is from an unmodified (UM) shock and the dashed
curve is the nonlinear (NL) result.
The shock smoothing and increase in $r$ produce substantial
differences in the spectra even though both shocks have exactly the
same input conditions.
(i) The overall normalization of the NL
spectrum is less, reflecting the conservation of energy
flux. 
(ii) The NL result has the distinctive concave curvature seen in
\nonrel\ shocks stemming from the fact that higher momentum particles
have a longer upstream diffusion length and get accelerated more
efficiently than lower momentum particles in the smooth shock.
(iii) The slope at the highest momentum in the NL spectrum reflects the
overall compression ratio and is flatter than the TP spectrum because
$r$ is greater.
(iv) The ``thermal'' peak is shifted to lower momentum in the NL
result and contains a larger fraction of mildly \rel\ particles than
in the UM result, i.e., $\Gamma \simeq 1.41$ for the NL shock
compared to $\Gamma \simeq 1.36$ for the UM shock.

\subsection{Fully relativistic, nonlinear shock: $\gamma_0 = 10$}

\begin{figure}[!hbtp]              
\epsscale{0.55} \plotone{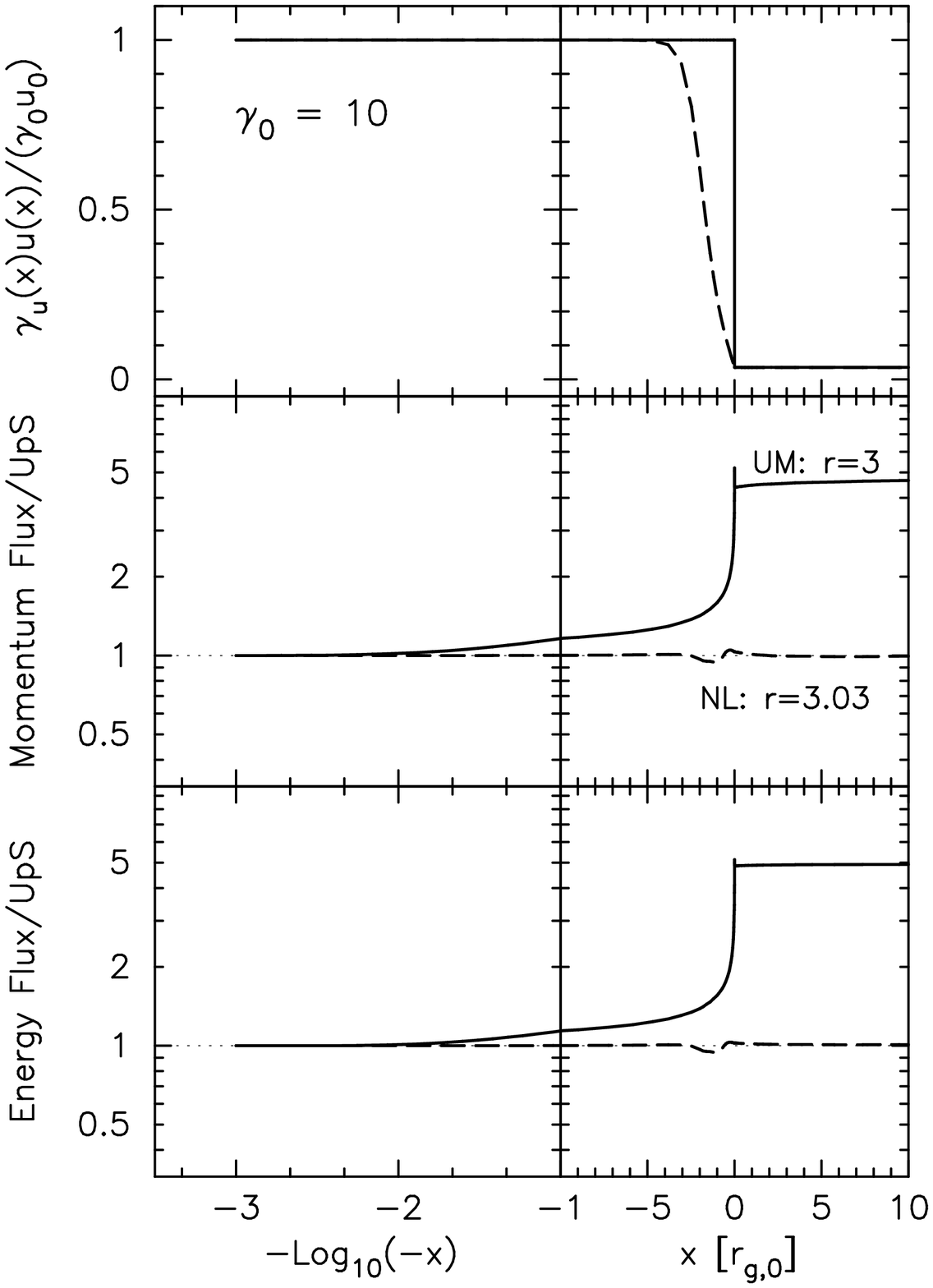} 
\figcaption{Unmodified (UM: solid curves) and nonlinear (NL: dashed
curves) shock profiles as in Fig.~\ref{fig:profG2} for
$\gamma_0=10$. The acceleration is truncated by $\pmax = 10^4 \beta_0
\gamsk m_p c$ and $N/\Nmin \sim 10$ in both cases.
\label{fig:profG10}
}
\end{figure}

\begin{figure}[!hbtp]              
\epsscale{0.55}
\plotone{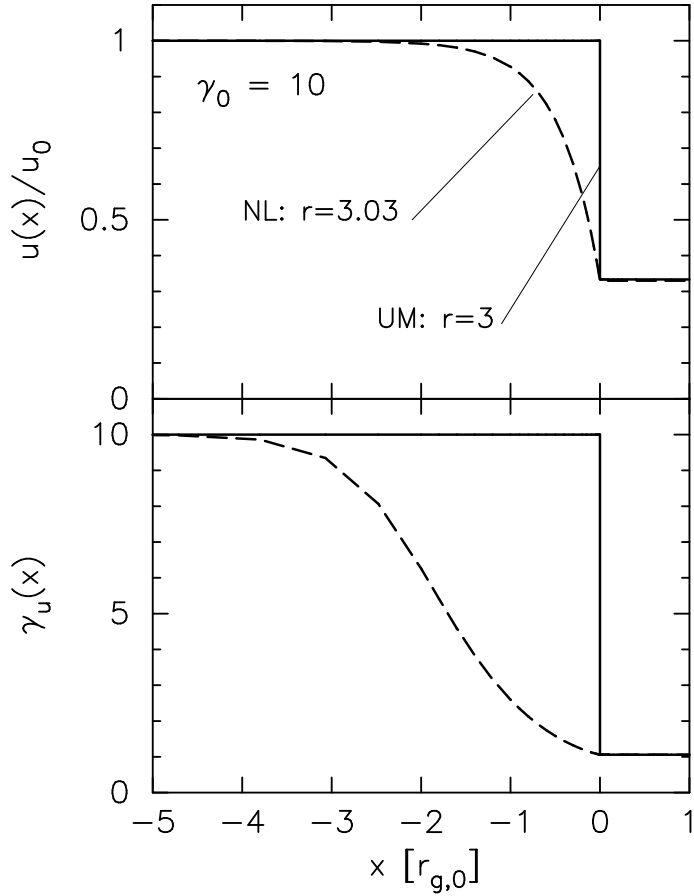}
\figcaption{The top panel is the flow speed at $x$ normalized to the
far upstream shock speed, $u_0$ versus $x$, for the $\gamsk=10$
shocks shown in Fig.~\ref{fig:profG10}. 
The bottom panel is the flow Lorentz factor,
$\gamma_u(x)$, versus $x$. In both panels, the solid curves are the
unmodified shock results with $r=3$ and the dashed curves are the
nonlinear results with $r\simeq 3.03 \pm 0.01$.
\label{fig:ugam}
}
\end{figure}

Fig.~\ref{fig:profG10} shows results for $\gamsk=10$.  In all
panels, the solid curves are for the unmodified shock, while the
dashed curves are for the flux conserving smoothed shock, both with
identical input parameters. As before, the acceleration is truncated
with a $\pmax = 10^4 \beta_0 \gamsk m_p c \simeq 9.3\xx{13}$ eV/c. In
Fig.~\ref{fig:ugam}, $u(x)$ and $\gamux$ are plotted separately.
Even though the length-scale of the shock smoothing is only a few
$\rgz$ and
the change in $r=3.03\pm 0.01$ is small, they
bring the momentum and energy fluxes into balance from values a factor
of five too large in the unmodified shock. The $\sim 1$\% difference
between $r=3.03\pm 0.01$ and the canonical value of 3 is significant; 
a self-consistent solution does not exist outside of this range.

As discussed above, both the shock structure and the overall
compression ratio may be modified when particle acceleration is
efficient. For mildly \rel\ shocks, such as our $\gamsk=1.4$ example,
a unique solution can be determined directly from the conserved
momentum and energy fluxes. For larger $\gamsk$'s however, the fluxes
are less sensitive to changes in the structure and compression ratio
and an additional constraint is needed to obtain a unique
solution. This constraint is provided by the \rel\ jump conditions 
for momentum and energy, i.e.,
\begin{equation}
\label{eq:mom_flx}
\gamma_{u0}^2 w_0 \frac{u_0^2}{c^2} + P_0 = 
\gamma_{u2}^2 w_2 \frac{u_2^2}{c^2} + P_2
\ ;
\end{equation}
\begin{equation}
\label{eq:en_flx}
\gamma_{u0}^2 w_0 u_0 = \gamma_{u2}^2 w_2 u_2
\ ,
\end{equation}
where $w = e + P$ is the enthalpy density, $e$ is the total energy
density, $P$ is the pressure, and we assume escaping fluxes are
negligible.  The energy density and pressure are related
through a combination of the adiabatic equation of state and the
conservation of energy,
i.e., 
\begin{equation}
P = (\Gamma -1)(e - \rho c^2)
\ ,
\end{equation}
where $\rho c^2$ is the rest mass energy density and $\Gamma$ 
is, in special cases, 
the ratio of specific heats \citep[e.g.,][]{ER91}.
Dividing Eq.~(\ref{eq:mom_flx}) by 
Eq.~(\ref{eq:en_flx}) yields
\begin{equation}
\label{vel_1}
\frac{u_0}{c^2} + \frac{P_0}{\gamma_{u0}^2(e_0 + P_0)u_0} =
\frac{u_2}{c^2} + \frac{P_2}{\gamma_{u2}^2(e_2 + P_2)u_2}
\ ,
\end{equation}
or, in terms of beta's,
\begin{equation}
\label{betas1}
\beta_0 + \left(\frac{P_0}{e_0 + P_0}\right)
\frac{1 - \beta_0^2}{\beta_0} =
\beta_2 + \left(\frac{P_2}{e_2 + P_2}\right)
\frac{1 - \beta_2^2}{\beta_2}
\ .
\end{equation}

The second term on the left hand side is small compared to
$\beta_0$ for unshocked upstream particle temperatures less than
$10^9$K
and upstream particle densities less than $100$ cm$^{-3}$. Neglecting
this term, Eq.~(\ref{betas1}) becomes
\begin{equation}
\label{betas2}
\beta_0 = 
\beta_2 + \left(\frac{P_2}{e_2 + P_2}\right)
\frac{1 - \beta_2^2}{\beta_2}
\ ,
\end{equation}
or,
\begin{equation}
\label{betaquad}
e_2\beta_2^2 - \beta_0(e_2 + P_2)\beta_2 + P_2 = 0
\ ,
\end{equation}
which has the shock solution
\begin{equation}
\label{betasol}
\beta_2 = \beta_0/r = 
\frac{1}{2e_2}
\left [ \beta_0(e_2 + P_2) - 
\sqrt{\beta_0^2(e_2 + P_2) - 4e_2P_2} \right ]
\ .
\end{equation}
For $\gamsk \gtrsim 10$, $\beta_0 \simeq 1$ and $r \simeq
e_2/P_2$. 
Furthermore, if \ultrarel\ downstream particle speeds can be assumed
so that $e_2 \gg \rho_2c^2$, Eq.~(\ref{betasol}) reduces to,
\begin{equation}
r \simeq \frac{1}{\Gamma_2 - 1}
\ .
\end{equation}
For $\Gamma_2=4/3$, $r \simeq 3$, the standard \ultrarel\ result.

\begin{figure}[!hbtp]              
\epsscale{0.55}
\plotone{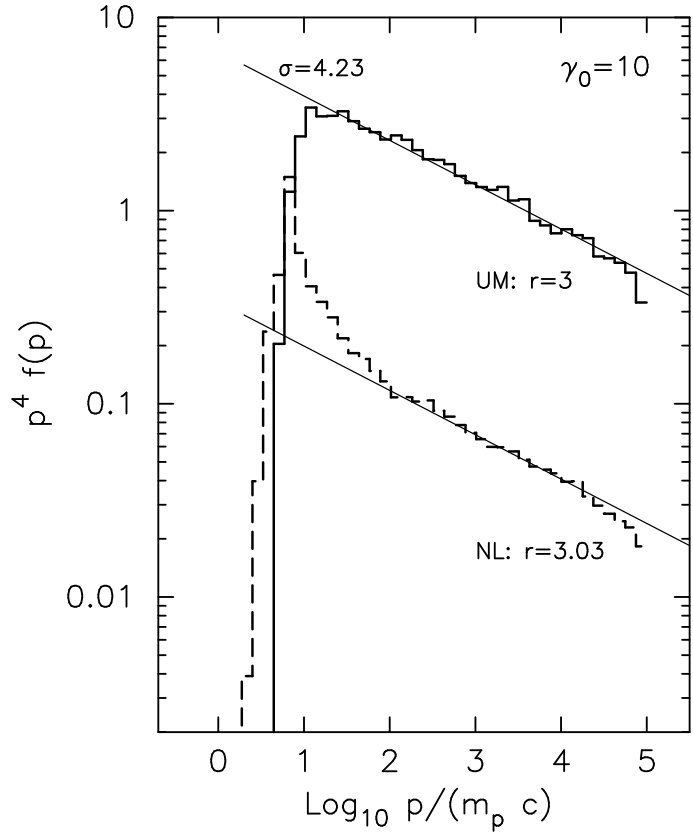} 
\figcaption{Particle distributions, $p^4 f(p)$, for the shocks shown
in Figs.~\ref{fig:profG10} and \ref{fig:ugam} with $\gamsk=10$. The
spectra for the nonlinear (NL) and unmodified (UM) shocks are
labeled and both are calculated at $x=0$ in the shock frame.
The light-weight solid lines show spectral indexes, $\sigma=4.23$.
\label{fig:specG10}
}
\end{figure}

To obtain a unique shock solution for $\gamsk \gtrsim 3$, we modify
both the shock structure and the compression ratio, $\rMC$, for the
Monte Carlo simulation, calculate $P_2$ and $e_2$ from the
resultant downstream particle distributions, and check that $\rMC
\simeq r$ as determined from Eq.~(\ref{betasol}). If $\rMC \ne
r$, the shock structure and $\rMC$ are varied until a consistent
solution is found with $\rMC \simeq r$.
For smaller $\gamsk$'s, Eq.~(\ref{betasol}) doesn't apply because
escaping fluxes become significant. In this case, however, the changes
in the momentum and energy fluxes from changes in the shock structure
and $\rMC$ are large enough that a unique solution can be found easily,
as in our $\gamsk=1.4$ example.

Despite the fact that $u(x)$ is modified on a fairly small
length scale, the resultant particle distribution function is changed
substantially, as indicated in Fig.~\ref{fig:specG10}.
In this figure, the solid curve is from the unmodified shock and the
dashed curve is from the nonlinear shock, both having exactly the same
input conditions. The unmodified spectrum in Fig.~\ref{fig:specG10} is
similar to that shown with a dashed curve in Fig.~\ref{fig:specTP}
only now 
the far upstream plasma is taken to be a thermal gas at a temperature
of $10^6$ K rather than a delta function distribution of particles
with speeds, $\vinj = (2 \Einj/m_p)^{1/2}$, with $\Einj= 1$ MeV, as
was assumed for the example in Fig.~\ref{fig:specTP}.
The shock smoothing has caused the low energy portion of the
distribution to steepen and the overall intensity to decrease,
reflecting the fact that the NL spectrum conserves momentum and energy
while the UM one doesn't.
The peaks in the two distributions at low momenta are also very
different, with the NL spectrum having a larger fraction of slower
particles than the UM one. These peaks result from the first shock
crossing where all particles receive a large energy gain.  In the UM
case, a far greater fraction of the accelerated downstream particles
are able to receive further energization by recrossing back into the
upstream region than in the NL shock.
The different speed distributions result in different $\Gamma$'s and
we find, by directly calculating $\Gamma$ from the distributions, that
$\GamTP \simeq 1.34$ while $\GamNL \simeq 1.36$.

\subsection{Acceleration efficiency}

\begin{figure}[!hbtp]              
\epsscale{0.55}
\plotone{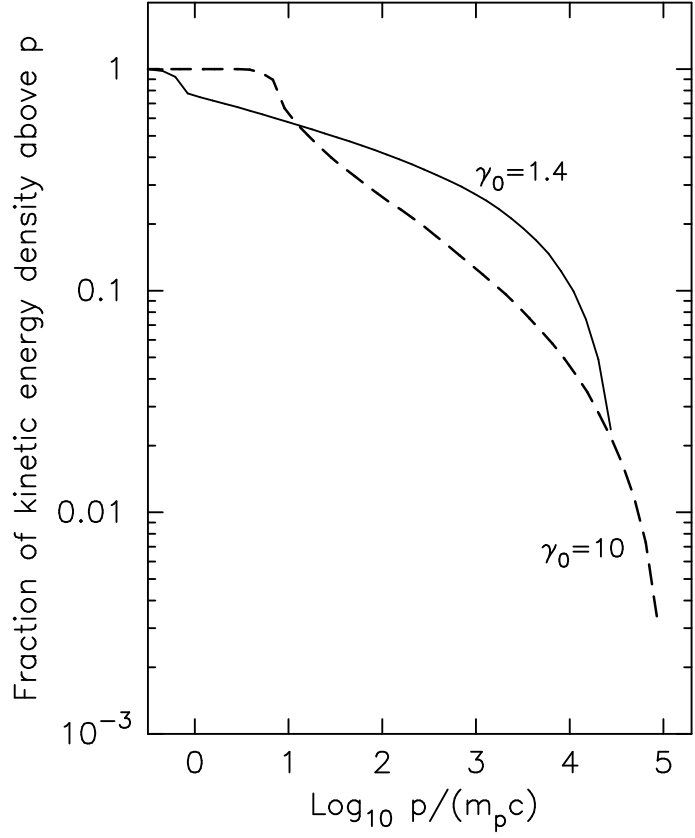}
\figcaption{Acceleration efficiency,
$\EffP$, defined as the fraction of total kinetic energy density above
$p$ versus $p$. The sharp drop off at low momentum indicates the extent
of the `thermal' peak.
The fraction of energy density above this drop off is a self-consistent
measure of the production efficiency for `superthermal' particles.
\label{fig:eff}
}
\end{figure}

The absolute acceleration efficiency can be determined in our
self-consistent, nonlinear examples directly from the particle
distributions. In Fig.~\ref{fig:eff} we plot, $\EffP$, i.e., the
fraction of kinetic energy density above a momentum $p$ versus $p$ for
our $\gamma_0=1.4$ and $\gamma_0=10$ examples. The fraction of kinetic
energy in the quasi-thermal part of the distribution can be determined
from the relatively sharp fall off of the distributions at low
momenta. This occurs near $m_p c$ for $\gamma_0=1.4$ and near $10\,
m_pc$ for $\gamma_0=10$. If we somewhat arbitrarily define the
acceleration efficiency for these two examples to be $\EffOneFour$ and
$\EffTen$, respectively, we have $\EffOneFour \simeq 0.7$ and $\EffTen
\simeq 0.6$. Of course, the behavior of $\EffP$ depends on the
particle spectrum from which it is derived and thus $\EffP$ is
flatter for the $\gamma_0=1.4$ shock, with $r\simeq 4.5$, than for the
$\gamma_0=10$ shock, where $r\simeq 3.03\pm 0.01$.  Furthermore,
$\EffP$ depends strongly on $\pmax$ in the $\gamsk=1.4$ shock where
particle escape plays an important role in determining
$r$. The maximum momentum has little influence for $\gamsk=10$ because
of the steep spectrum.
The fraction of kinetic energy density above $10^3 m_p c$ is $\sim 0.3$ for
$\gamsk=1.4$ and $\sim 0.1$ for $\gamsk=10$.

\section{Summary and Conclusions}

Particles gain energy in collisionless shocks by scattering nearly
elastically off magnetic turbulence, back and forth between the
converging plasmas upstream and downstream from the shock.
While this basic shock acceleration physics is independent of the
speed of the shock, the mathematical modeling of the process depends
critically on whether or not the acceleration is efficient and whether
or not particle speeds, $v$, are large compared to the shock speed,
$u_0$.
The Monte Carlo techniques discussed here, which do not require $v \gg
u_0$, are well suited for the study of \rel\ shocks, and for any shock
where nonlinear effects are important and the energetic particles
originate as thermal particles in the unshocked plasma.  Except for
computational limits, these techniques allow calculations of efficient
particle acceleration in shocks of any Lorentz factor.

As a check of our code, we have demonstrated that we obtain the well
known, \TP\ power laws in fully \nonrel\ and \ultrarel\ parallel
shocks (Figs.~\ref{fig:gyro} and \ref{fig:specTP}).  In \transrel\
shocks, however, no such canonical results exist because the shock
compression ratio, $r$, depends on the upstream conditions in a
nonlinear fashion, even for \TP\ shocks where all effects of
accelerated particles are ignored \citep[e.g.,][]{ER91}.  We show how
$r$, and the resulting power law spectral index, $\sigma$, vary
through the \transrel\ regime in Fig.~\ref{fig:ratio}, where $r$ has
been determined by balancing the momentum and energy fluxes across the
shock in an iterative process.
These \TP\ results with self-consistently determined compression
ratios and adiabatic indexes are in close agreement with the recent
analytic results of \citet{KGGA2000} and \citet{Gallant2002}.

Despite the fact that relativistic shock theory has concentrated
almost exclusively on \TP\ acceleration, it is likely that \rel\
shocks are not \TP\ but inject and accelerate particles
efficiently. The reason is that regardless of the ambient far upstream
conditions, all particles that are over taken by an \ultrarel\ shock
will receive a large boosts in energy $\Delta E > \gamsk m c^2$ in
their first shock crossing. Thus, virtually all of the particles in
the downstream region of an unmodified shock are strongly \rel\ with
$v\sim c$.
The ability to overtake the shock from downstream and be further
accelerated depends only on the particle speed
(Eq.~\ref{eq:ProbRet}) and the presence of magnetic waves or
turbulence with sufficient power in wavelengths on the order of the
particle gyroradii to isotropize the downstream distributions. 
It is generally assumed that the necessary magnetic turbulence is
self-generated and if enough turbulence is generated to scatter high
momentum particles (with very low densities) that constitute a \TP\
power law, there should be enough generated to isotropize lower
momentum particles (which carry the bulk of the density). If
acceleration can occur at all, we believe it is likely to occur
efficiently making it necessary to calculate the shock structure and
particle acceleration self-consistently.
Furthermore, if \rel\ shock theory is to be applied to gamma-ray
bursts, where high conversion efficiencies are generally assumed,
nonlinear effects must be calculated.

When energetic particles are generated in sufficient numbers, the
conservation of momentum and energy requires that their backpressure
modify the shock structure. Two basic effects occur: a precursor
is formed when the upstream plasma is slowed by the backpressure of
the accelerated particles and the overall compression ratio changes
from the \TP\ value as a result of high energy particles escaping
and/or a change in the shocked plasma's 
adiabatic index
$\Gamma$.
As indicated by our $\gamsk=1.4$ example (Section 4.1), mildly \rel\
shocks act as \nonrel\ ones showing a dramatic weakening of the
subshock combined with a large increase in $r$
(Fig.~\ref{fig:profG2}). These changes result in a particle
distribution which is both steeper than the \TP\ power law at low
momenta and flatter at high momenta (Fig.~\ref{fig:specG2}).

In faster shocks (i.e., $\gamsk \gtrsim 3$), the initial \TP\ spectrum
is steep enough that particle escape is unimportant so only changes in
$\Gamma$ determine $r$ (Eq.~\ref{betasol}). In contrast to
\nonrel\ shocks where the production of \rel\ particles causes the
compression ratio to increase, we show that $r$ decreases smoothly to
$3$ as $\gamsk$ increases and the fraction of fully \rel\ shocked
particles approaches one (Fig.~\ref{fig:ratio}).

Our most important result is that efficient, mildly \rel\ shocks do
not produce particle spectra close to the so-called `universal' power
law having $\sigma \sim 4.3$. This may be important for the
interpretation of gamma-ray bursts since the internal shocks assumed
responsible for converting the bulk kinetic energy of the fireball
into internal particle energy may be mildly \rel\ and the
external shocks, believed responsible for producing gamma-ray burst
afterglows, will always go through a mildly \rel\ phase
\citep[see][for a comprehensive review of gamma-ray bursts]{Piran99}.

\acknowledgments The authors thank M. G. Baring, L. O'C. Drury,
F. C. Jones, and E. Parizot for helpful
discussions. D.C.E. is grateful to the Department of Physics and
Astronomy at Rice University and to the Dublin Institute for Advanced
Studies for hosting visits where part of this work was done.

\newcommand{\aaDE}[3]{ 19#1, A\&A, #2, #3}
\newcommand{\aatwoDE}[3]{ 20#1, A\&A, #2, #3}
\newcommand{\aasupDE}[3]{ 19#1, {\itt A\&AS,} {\bff #2}, #3}
\newcommand{\ajDE}[3]{ 19#1, {\itt AJ,} {\bff #2}, #3}
\newcommand{\anngeophysDE}[3]{ 19#1, {\itt Ann. Geophys.,} {\bff #2}, #3}
\newcommand{\anngeophysicDE}[3]{ 19#1, {\itt Ann. Geophysicae,} {\bff #2}, #3}
\newcommand{\annrevDE}[3]{ 19#1, {\itt Ann. Rev. Astr. Ap.,} {\bff #2}, #3}
\newcommand{\apjDE}[3]{ 19#1, {\itt ApJ,} {\bff #2}, #3}
\newcommand{\apjtwoDE}[3]{ 20#1, {\itt ApJ,} {\bff #2}, #3}
\newcommand{\apjletDE}[3]{ 19#1, {\itt ApJ,} {\bff  #2}, #3}
\newcommand{\apjlettwoDE}[3]{ 20#1, {\itt ApJ,} {\bff  #2}, #3}
\newcommand{\apjpress}{{\itt ApJ,} in press}
\newcommand{\apjletpress}{{\itt ApJ(Letts),} in press}
\newcommand{\apjsDE}[3]{ 19#1, {\itt ApJS,} {\bff #2}, #3}
\newcommand{\apjsubDE}[1]{ 19#1, {\itt ApJ}, submitted.}
\newcommand{\apjsubtwoDE}[1]{ 20#1, {\itt ApJ}, submitted.}
\newcommand{\appDE}[3]{ 19#1, {\itt Astroparticle Phys.,} {\bff #2}, #3}
\newcommand{\apptwoDE}[3]{ 20#1, {\itt Astroparticle Phys.,} {\bff #2}, #3}
\newcommand{\araaDE}[3]{ 19#1, {\itt ARA\&A,} {\bff #2},
   #3}
\newcommand{\assDE}[3]{ 19#1, {\itt Astr. Sp. Sci.,} {\bff #2}, #3}
\newcommand{\icrcplovdiv}[2]{ 1977, in {\itt Proc. 15th ICRC(Plovdiv)},
   {\bff #1}, #2}
\newcommand{\icrcsaltlake}[2]{ 1999, {\itt Proc. 26th Int. Cosmic Ray Conf.
    (Salt Lake City),} {\bff #1}, #2}
\newcommand{\icrcsaltlakepress}[2]{ 19#1, {\itt Proc. 26th Int. Cosmic Ray Conf.
    (Salt Lake City),} paper #2}
\newcommand{\jgrDE}[3]{ 19#1, {\itt J.G.R., } {\bff #2}, #3}
\newcommand{\mnrasDE}[3]{ 19#1, {\itt M.N.R.A.S.,} {\bff #2}, #3}
\newcommand{\mnrastwoDE}[3]{ 20#1, {\itt M.N.R.A.S.,} {\bff #2}, #3}
\newcommand{\mnraspress}[1]{ 20#1, {\itt M.N.R.A.S.,} in press}
\newcommand{\natureDE}[3]{ 19#1, {\itt Nature,} {\bff #2}, #3}
\newcommand{\pfDE}[3]{ 19#1, {\itt Phys. Fluids,} {\bff #2}, #3}
\newcommand{\phyreptsDE}[3]{ 19#1, {\itt Phys. Repts.,} {\bff #2}, #3}
\newcommand{\physrevEDE}[3]{ 19#1, {\it Phys. Rev. E,} {\bf #2}, #3}
\newcommand{\prlDE}[3]{ 19#1, {\it Phys. Rev. Letts,} {\bf #2}, #3}
\newcommand{\revgeospphyDE}[3]{ 19#1, {\itt Rev. Geophys and Sp. Phys.,}
   {\bff #2}, #3}
\newcommand{\rppDE}[3]{ 19#1, {\itt Rep. Prog. Phys.,} {\bff #2}, #3}
\newcommand{\ssrDE}[3]{ 19#1, {\itt Space Sci. Rev.,} {\bff #2}, #3}

\end{document}